\documentstyle[12pt]{article}
\font\tenrm=cmr10

\def\ga{\gamma}
\def\de{\delta}
\def\ep{\epsilon}
\def\ve{\varepsilon}

\def\la{\lambda}

\def\ph{\phi}

\def\ch{\chi}
\def\ps{\psi}

\def\Ga{\Gamma}
\def\De{\Delta}

\def\Ph{\Phi}

\def\cl{{\cal L}}
\def\fr#1#2{{{#1} \over {#2}}}
\def\prt{\partial}
\def\vev#1{\langle {#1}\rangle}
\def\lsim{\mathrel{\rlap{\lower4pt\hbox{\hskip1pt$\sim$}}
    \raise1pt\hbox{$<$}}}
\def\gsim{\mathrel{\rlap{\lower4pt\hbox{\hskip1pt$\sim$}}
    \raise1pt\hbox{$>$}}}

\def\ket#1{|{#1}\rangle}

\def\half{{\textstyle{1\over 2}}}
\def\frac#1#2{{\textstyle{{#1}\over {#2}}}}

\newcommand{\beq}{\begin{equation}}
\newcommand{\eeq}{\end{equation}}
\newcommand{\bea}{\begin{eqnarray}}
\newcommand{\eea}{\end{eqnarray}}
\newcommand{\rf}[1]{(\ref{#1})}

\renewenvironment{thebibliography}[1]
 { \rm
   \begin{list}{\arabic{enumi}.}
    {\usecounter{enumi} \setlength{\parsep}{0pt}
     \setlength{\itemsep}{3pt} \settowidth{\labelwidth}{#1.}
     \sloppy
    }}{\end{list}}

\begin{document}

\begin{flushright}
{IUHET 236\\}
{hep-ph/9211116\\}
{October 1992\\}
\end{flushright}

\begin{center}
{{\bf CPT, STRINGS, AND THE $K \bar K$ SYSTEM\footnote{
\tenrm Invited talk given by V.A.K. at the Los Angeles meeting
on Low Energy Signals from the Planck Scale, October 29-30 1992}\\}
\vglue 1.0cm
{V. ALAN KOSTELECK\'Y\\}
\bigskip
{\it Physics Department, Indiana University,\\}
\medskip
{\it Bloomington, IN 47405, U.S.A.\\}
\vglue 0.3cm
{and\\}
\vglue 0.3cm
{ROBERTUS POTTING\\}
\bigskip
{\it U.C.E.H., Universidade do Algarve, Campus de Gambelas,\\}
\medskip
{\it 8000 Faro, Portugal\\}

\vglue 0.8cm
{\tenrm ABSTRACT}}
\vglue 0.3cm

\end{center}

{\rightskip=3pc\leftskip=3pc\tenrm\noindent
This talk contains a summary of our work on dynamical CPT invariance
and spontaneous CPT violation in string theories, including the
possibility that stringy CPT violation could
occur at levels detectable in the next generation of experiments.
In particular,
we present here an estimate for values of parameters for CPT violation
in the kaon system.

}

\baselineskip=14pt
\vglue 0.6cm
{\bf\noindent 1. Introduction}
\vglue 0.4cm

String theory is an ambitious proposal for a consistent quantum theory of
gravity, simultaneously providing a framework for unification of the
fundamental forces and interactions.
Like any new physical model,
string theory faces two issues.
First, is it viable?
Second, is it falsifiable?
Viability includes the issues of
internal self-consistency and physical realism.
Falsifiability is satisfied if new effects are predicted
that are experimentally observable.

Most of the work on string theory
has addressed the question of viability,
in particular the issue of realism.
In contrast, relatively little has been done on the difficult issue of
falsifiability,
i.e.,
the existence and observability of stringy effects.
A stringy effect can be taken as one violating
some principle accepted as valid for a pure particle theory.
Such effects are to be expected because strings are
qualitatively different from particles,
but there is a folklore that they are unobservable in current experiments
because a low-energy string theory looks like a point-particle theory.
However,
it is at least possible that
a string-based low-energy effective theory
is only approximately
a standard, four-dimensional, renormalizable gauge field theory.

Proving or disproving the existence of such effects
in general is evidently difficult
in the absence of a completely realistic string theory.
However, it is feasible to examine mechanisms
by which the string \it might \rm signal itself to us at low energies,
without seeking to address the question of whether they \it must \rm occur.
Even this lesser task is daunting
because any such effects should be highly suppressed.
Present experiments can at best attain energies of the order of
the electroweak scale $m_W$.
This is only
about $10^{-17}$ of the natural string scale,
the Planck mass $m_P$.
To minimize the impact of this large suppression factor,
it is reasonable to focus on physical features that are believed to
be exact and generic in particle theory
and that can be tested to a high degree of precision.

This talk presents our work on one possible
string signature, CPT violation.
It outlines our published work \cite{KP} on dynamical CPT invariance
and spontaneous CPT violation in string theories.
In addition,
we provide here more detail on the connection between
stringy CPT violation and experimentally observable parameters
in the $K \bar K$ system.

\vglue 0.6cm
{\bf\noindent 2. CPT and Strings}
\vglue 0.4cm

The CPT theorem \cite{S} is a profound result
holding for local relativistic point-particle field theories
under a few mild restrictions.
The theorem has substantial experimental support:
at present,
the tightest bound is found from
observations of the $K \bar K$ system,
where one figure of merit is \cite{C}
\beq
\fr {m_K- m_{\bar K}} {m_K} \le 5\times 10^{-18}
\quad .
\label{a}
\eeq
The generality of the theoretical result for particles,
together with the extended nature of strings
and the existence of a high-precision test,
make CPT a good candidate for a possible string signature.

Given a field theory,
either for particles or for strings,
several methods to investigate CPT properties are in principle available.
The axiomatic method is perhaps the most general,
but suitable extensions of the Wightman axioms for strings have not yet
been formulated.
An alternative is the constructive method,
in which C, P, T are defined on anticommuting irreducible spinors
and the CPT transformation properties of arbitrary irreducible
tensor products are inferred.
However,
this would require more information on the general structure of covariant
string theories.
Instead,
we use a practical method
\cite{SA} that can be applied on a case-by-case basis.
First, one seeks C, P, T transformations for the
free theory that are consistent with usual notions
and with free-field quantization.
Assuming this is possible,
one then examines the effect of the C, P, T transformations
on the interacting fields by expanding them in free fields
in the usual perturbation series.
If the interactions change the canonical structure,
which happens for strings,
the consistency of the transformations under the interacting-field
quantization procedure must be verified.
Additional constraints may arise if certain
fields in the theory are to be identified with physical particles
that have known transformation laws.
In what follows,
we take the massless modes of the string to transform
like the particle fields in the standard model.

There are two issues to examine when considering
CPT properties of a theory.
One, dynamical CPT, is the behavior of the action
under the transformation.
The other is the possibility of spontaneous CPT violation,
which occurs if the ground state is not invariant under CPT.
Subsection 3 summarizes our results on dynamical CPT invariance,
while subsection 4 contains a discussion of the possibility
of spontaneous CPT violation.

\vglue 0.6cm
{\bf \noindent 3. Dynamical CPT Invariance}
\vglue 0.4cm

To investigate the issue of dynamical CPT,
we studied
the field theory for the open bosonic string \cite{W}
and those proposed for the open superstring \cite{PTY,AMZ}.
As the string theory contains an infinite number of particle modes,
the statement of dynamical CPT invariance implies the assignment
of an infinite number of CPT transformation rules
that are consistent with the free case,
with the infinite number of interactions,
and with the interacting-field quantization.
Fortunately,
the formalism of string field theory permits a direct treatment
in terms of string fields.
The CPT transformations of the modes are important as such,
however,
because experiments detect particles, not strings.
Also, even though at present only the lowest string modes
would be accessible,
higher modes might nonetheless play a significant role
because they are incorporated into the low-energy effective theory
via functional integration.
It is therefore useful to construct operators that
implement C, P, and T at the mode level.
This construction is also presented in \cite{KP}.

Here is a summary of our results.
The action of the open bosonic string
is invariant under C, P, T and CPT.
All the actions proposed for the open superstring
violate C, P, and T but preserve CPT.
While the P and T violations are expected for a ten-dimensional
theory that contains massless chiral fermions,
the C violation is not.
However,
we were able to find a modified action for the open superstring
that preserves C and CPT.

In any event,
it seems that the differences
between particle and string quantum field theories
are insufficient to cause dynamical CPT violation.
The extended structure and corresponding smeared particle interactions
are apparently smooth enough to preserve CPT,
while the presence of an infinite number of fields does not cause
trouble because each field enters as a finite representation
of the Lorentz group.
Our analysis makes use of some standard assumptions,
such as the validity of the perturbation series,
the completeness of the asymptotic Hilbert space,
the reality of observables,
and the connection between spin and statistics.
As such,
it is likely to extend to field theories for other string models.

\vglue 0.6cm
{\bf \noindent 4. Spontaneous CPT Violation and the $K \bar K$ System}
\vglue 0.4cm

The remaining issue is whether
the vacuum is CPT invariant.
It is known \cite{KS}
that there is a natural mechanism in string theory
for the spontaneous breaking of Lorentz invariance.
This mechanism can also break CPT.

To see how the mechanism might work,
consider the open bosonic string.
Among the interaction terms of this theory
is a coupling $\phi A_\mu A^\mu$
between the tachyon $\phi$ and the massless vector $A_\mu$.
The (static) effective potential for $\phi$ has an instability
at the origin,
suggesting the tachyon acquires a vacuum expectation value (vev).
If this vev has the appropriate sign,
the trilinear coupling above
generates a negative contribution to the squared mass
of the vector.
This destabilizes in turn the effective potential for the vector,
which could cause it to acquire a nonzero vev and hence to
spontaneously break the (higher-dimensional) Lorentz symmetry.
Note that this procedure cannot happen in a standard four-dimensional
gauge field theory:
particle gauge invariance excludes trilinear couplings of the necessary form,
even though they are compatible with string gauge invariance.
The mechanism can in principle involve any Lorentz tensor in a string model,
because such tensors always have trilinear couplings to scalars.
See ref. \cite{KS} for more details.

If the mechanism occurs,
it can also break CPT.
For instance,
in the example above the vector $A_\mu$ changes sign under CPT,
so, if it acquires a vev, terms breaking CPT appear in the lagrangian.
More generally,
if CPT induces a sign change in a field $f$
and if a vev is generated for $f$,
then any interaction terms involving $f$
generate CPT violation when the shift to the new (stable) vacuum is made.
This could in principle include violation of four-dimensional CPT.

Evidently,
any such effects must be suppressed in the low-energy four-dimensional
effective theory,
perhaps because only the higher (Planck-mass) modes are involved.
In any case,
the natural suppression factor is the ratio of the low-energy scale $m_l$
to the Planck scale $m_P$,
which is about $10^{-17}$ or less.
For example,
in the $K \bar K$ system one might expect \cite{KP}
\beq
\fr {m_K- m_{\bar K}} {m_K} \sim \fr {m_l}{m_P}
\quad ,
\label{b}
\eeq
which is in the range $10^{-20}$ to $10^{-17}$.

To make this more explicit,
consider a class of
four-dimensional low-energy effective interactions
from string theory:
\beq
\cl_I \supset \la m_P^{-k} T \cdot \bar\ps \Ga (i\prt )^k \ch
\quad .
\label{c}
\eeq
In this equation,
$\la$ is a dimensionless coupling,
the factor $m_P^{-k}$ is a dimensional factor
correcting for derivative couplings and the compactification process,
$T$ is a four-dimensional Lorentz tensor,
$\ps$ and $\ch$ are four-dimensional fermions (possibly the same),
$\Ga$ represents a gamma-matrix structure,
and $(i\prt )^k$ denotes the presence of $k$ derivatives $\prt_\mu$.
Since $\cl_I$ is a Lorentz scalar,
$T \cdot \Ga (i\prt )^k$ is a spinor matrix with derivative entries.

If $T$ acquires a vev $\vev T$,
$\cl_I$ generates a contribution $\De K(p)$ to the
fermion inverse propagator $K(p)$,
given by
\beq
\De K(p) = \la m_P^{-k} \vev T \cdot \Ga p^k \quad  ,\qquad  k \ge 0
\quad .
\label{d}
\eeq
Suppose the vev $\vev T$ is written as
\beq
\vev T = t \left( \fr {m_l}{m_P} \right)^l m_P \quad ,\qquad l \ge 0
\quad ,
\label{e}
\eeq
where $t$ is a numerical factor that may carry Lorentz indices
and the factor $(m_l/m_P)^l$ allows for a suppression factor
in line with the known hierarchy in nature.
Then,
\beq
\fr{\De K}{K} \sim \left( \fr {p}{m_P} \right)^k
\left( \fr {m_l}{m_P} \right)^{l-1}
\left( \fr {m_l}{m_f} \right)
\quad ,
\label{f}
\eeq
where $m_f$ is the fermion mass.
If the fermions are taken to be light (observable)
then $p \ll m_P$ and so $k+l \ge 1$.
Moreover,
if $T$ has nontrivial Lorentz structure
and incorporates CPT breaking,
a realistic model presumably has $\vev T \ll m_l$, i.e., $l \ge 2$.
Then, the issue of whether any stringy spontaneous CPT violation
might be accessible to experiment becomes the question of
whether terms with $k=0$ and $l=2$ are observable.

Since such terms are suppressed by a factor of $10^{-17}$ or more,
we need a sensitive (interferometric) experiment
to observe their effects.
A candidate is the $K \bar K$ system.
For this case,
take the fermions $\ps$ and $\ch$ to be $d$ and/or $s$ quarks.
One can work in lowest-order perturbation theory where necessary,
replacing $\De K(p)$ with its expectation in the quark wavefunction.
For simplicity,
suppose the expectation value $\vev T = t (m_l/m_P)^2 m_P$
and the gamma-matrix structure $\Ga$ are such that the effect
on the $d$ or $s$ inverse propagators is an energy shift.
This happens, for instance, for interactions of the form
$T_{\la\mu\nu}\bar\ps\ga^\la\ga^\mu\ga^\nu\ch$
if $T$ acquires a vev $\vev {T_{000}}$.
The energy shifts for the $d$ and $s$ quarks induce energy changes in the
$K$ and $\bar K$,
thereby affecting the kaon mass/decay matrix.

Before proceeding further,
we summarize some notation and conventions.
We take the states
$\ket{K^0}$ and $\ket{\bar K^0}$
to satisfy
\bea
CP \ket{K^0} & = & -\ket{\bar K^0} \\
CP \ket{\bar K^0} & = & -\ket{K^0}
\quad ,
\label{g}
\eea
The effective hamiltonian governing the time evolution of
a superposition of
$\ket{K^0}$ and $\ket{\bar K^0}$
is a two-by-two matrix $H$:
\beq
H=M-i\Ga \equiv
\pmatrix{H_{11}&H_{12}\cr
H_{21}&H_{22}\cr}
\quad .
\label{h}
\eeq
Under CPT, $H_{11} \to H_{22}$ and vice versa.
A useful parametrization is
\beq
iH=\Ga -iM \equiv
\pmatrix{D+iE_3&iE_1+E_2\cr
iE_1-E_2&D-iE_3\cr}
\quad .
\label{i}
\eeq

The eigenvalues $\la_L$ and $\la_S$ of this matrix and the
corresponding eigenvectors
$\ket{K_L}$ and $\ket{K_S}$
are given by
\bea
\la_L & = & D+iE\quad ,\qquad
\ket{K_L}= \fr {(1+\ep - \de ) \ket{K^0} +(1-\ep +\de )\ket{\bar K^0}}
{\sqrt{2(1+\vert \ep - \de \vert^2)}}
\quad , \\
\la_S & = &D-iE\quad ,\qquad
\ket{K_L}= \fr {(1+\ep + \de ) \ket{K^0} -(1-\ep -\de )\ket{\bar K^0}}
{\sqrt{2(1+\vert \ep + \de \vert^2)}}
\quad .
\label{j}
\eea
In these equations,
\beq
E^2 = E_1^2+E_2^2+E_3^2
\quad ,
\label{k}
\eeq
while
\beq
\ep = \fr { -iE_2}{E_1 + E}
\quad
\label{l}
\eeq
measures T violation and
\beq
\de = \fr { -E_3}{E_1 + E}
\quad
\label{m}
\eeq
measures CPT violation.

Introducing the masses and lifetimes
of the $K_L$ and the $K_S$ via
\beq
\la_L = \half \ga_L + im_L\quad ,\qquad \la_S = \half \ga_S + im_S
\quad ,
\label{n}
\eeq
we get
\beq
E = \half \De m + \frac 1 4  i \De \ga
\quad ,
\label{o}
\eeq
where $\De \ga = \ga_S - \ga_L$ and $\De m = m_L - m_S$.
Experimentally, $\De m \simeq \half\De \ga$,
and one finds $E_1 \simeq E$,
so defining the superweak phase
\beq
\ph_\ep = \tan^{-1} \fr {2\De m}{\De\ga}
\quad
\label{p}
\eeq
gives
\beq
\de \simeq \fr i {\sqrt 2} \fr {E_3}{\De m} e^{+i\ph_\ve}
\quad .
\label{q}
\eeq
The quantities $\De m =(3.522\pm 0.016)\times 10^{-15}$GeV
and $\ph_\ep = 43.68 \pm 0.14^\circ$
are experimentally determined \cite{PDG}.

This formalism can be used to obtain an expression for
the CPT-violating parameter $\de$ in terms of the
quantities arising in the low-energy effective interactions.
The energy shifts for the $d$ and $s$ quarks generate
corresponding shifts in the kaon matrix $H$:
\beq
\De H_{11} = - \De H_{22} = h_d - h_s
\quad
\label{r}
\eeq
where
\beq
h_{d,s} = r_{d,s} \la_{d,s} t \fr {m_l^2}{m_P}
\quad .
\label{s}
\eeq
The extra factors $r_{d,s}$
are QCD corrections allowing for quark-binding effects.
Setting $\De H_{11} = E_3$ gives
\beq
\de \simeq \fr i {\sqrt 2} \fr {h_d-h_s}{\De m} e^{+i\ph_\ve}
\quad .
\label{t}
\eeq

The issue of observability of possible CPT-violating effects
then reduces to estimating the net effect of contributions
to $\de$ of the form \rf{t} from different interactions.
If the assumption is made that the energy shifts are real,
we see that Re$~\de \simeq -$Im$~\de$.
Then, for $m_l$ lying in the region around $m_K$ to $m_W$,
it is possible to get contributions to $\vert$Re$~\de\vert$
lying below the current experimental limit of about $10^{-4}$
down to about $10^{-6}$.
Part of this region may be explored by the $\Ph$ factories
under development at Frascati and Novosibirsk \cite{T}
or at the novel asymmetric $\Ph$ factory being considered
at UCLA \cite{CL}.

In principle,
CPT violating effects other than those included in $\de$ might arise
in the decay amplitudes \cite{TD}.
However, the energy shifts discussed above
do not give a direct contribution to the matrix elements
of the decays.
There is an indirect effect that could arise through the
energy-dependent normalizations of the eigenstates,
but it is suppressed by at least one power of
$m_l/m_P$
and so is negligible.
Another contribution comes from
interactions of the form \rf{c}
when $\ps$ is $d$ and $\ch$ is $s$ (and vice versa).
However,
these either give negligible additional contributions
to electroweak-type flavor-changing processes,
or they generate standard-model-forbidden decays
at unobservable levels because a suppression of at least two powers of
$m_l/m_P$ appears.

\vglue 0.6cm
{\bf \noindent 5. Summary}
\vglue 0.4cm

In summary,
CPT is a good candidate for a string signature.
We have demonstrated dynamical CPT invariance
in field theories for the open bosonic string and the open superstring.
A natural mechanism exists in string theory that could lead to spontaneous
CPT breaking.
It can induce violation in the $K \bar K$ system
at levels just beyond current limits.
The signal would be a finite value for $\de$
and negligible other CPT-violating parameters.

\vglue 0.6cm
{\bf \noindent 6. Acknowledgments}
\vglue 0.4cm
We thank David Cline, Stuart Samuel, Julia Thompson,
and Mike Zeller for useful conversations.
V.A.K. thanks the Theory Division of CERN and the Aspen Center for Physics
for hospitality while part of this work was done.
This work was supported in part by the United States Department of
Energy under contracts DE-AC02-84ER41025 and DE-FG02-91ER40661.

\vglue 0.6cm
{\bf\noindent 7. References}
\vglue 0.4cm

\end{document}